\documentclass[preprintnumbers, pra, showpacs, floatfix,twocolumn,
preprintnumbers, letterpaper, superscriptaddress]{revtex4-2}
\usepackage{amsfonts}
\usepackage{amsmath}
\usepackage{amssymb,epsf}
\usepackage{latexsym}
\usepackage{graphicx,epsfig}
\usepackage{amssymb}
\usepackage{subfigure}
\usepackage[T1]{fontenc}
\usepackage[utf8]{inputenc}
\usepackage[colorlinks=true,citecolor=blue,linkcolor=blue,urlcolor=black]{hyperref}
\usepackage{epstopdf}
\usepackage{color}
\usepackage{physics}
\usepackage{comment}

\begin{document}
\title{Spin stiffness and resilience phase transition in a noisy toric-rotor code}
\author{Morteza Zarei}
\email{m.zarei@shirazu.ac.ir}
\affiliation{Physics Department, College of Sciences, Shiraz University, Shiraz 71454, Iran}
\author{Mohammad Hossein Zarei}
\email{mzarei92@shirazu.ac.ir}
\affiliation{Physics Department, College of Sciences, Shiraz University, Shiraz 71454, Iran}
\begin{abstract}
We use a quantum formalism for the partition function of the classical $XY$ model to identify a resilience phase transition in the zero-syndrome postselected sector of a noisy toric-rotor code. To this end, we consider a logical state of toric-rotor code under phase-shift noise described by a von Mises probability distribution. We then show that the fidelity of the noisy state with respect to the initial logical state is proportional to the partition function of the $XY$ model, such that a Kosterlitz-Thouless phase transition at a critical temperature $T_c$ corresponds to a resilience phase transition at a critical width $\sigma_c$. To characterize this transition, we map the spin stiffness of the $XY$ model to a topological order parameter $0\leq \lambda \leq 1$, which quantifies the intrinsic resilience of the code to decoherence within the zero-syndrome subspace. We show that the initial logical state exhibits partial resilience to noise for widths less than $\sigma_c \approx 0.89$, where $\lambda$ satisfies $0< \lambda <1$ and drops
discontinuously to zero at $\sigma_c$. We further discuss the implications of our results for postselected quantum error correction in the toric-rotor code in higher dimensions. Our work shows that the quantum formalism for partition functions provides a mathematically rigorous framework for studying noisy continuous-variable quantum codes.	
\end{abstract}

\maketitle
\section{Introduction}
Quantum error-correcting codes are among the best-known approaches for overcoming noise in quantum computation \cite{Shor1995,Steane1996,Gottesman1997}. In particular, the interplay between quantum error correction and quantum topological order has attracted significant attention because of the natural robustness of topological phases to local perturbations \cite{Kitaev2003,Chen2010,Wen2017,Zarei2015b}. In topological quantum codes, logical information is stored in a global, topological property of the system in the sense that local errors cannot damage this global information \cite{Nayak2008,Terhal2015}.

The toric code is one of the best-known topological codes and has been extensively studied over the past two decades \cite{Fowler2012c,Jamadagni2018,Zhu2022,Oliveira2024,Cong2024}. The intrinsic robustness of topological order in the toric code is reflected in its finite fault-tolerant threshold \cite{Dennis2002,Bombin2012c} when the system is subjected to a noisy environment. This threshold phenomenon is, in fact, related to a phase transition in the noisy toric code from a correctable to a non-correctable phase \cite{Wang2003a,Vodola2022,Zhao2024}. Interestingly, the above phase transition is mapped to a classical phase transition in the random-bond Ising model (RBIM).

Studying such phase transitions plays an important role in understanding topological order in noisy topological quantum codes \cite{Zhao2024,Sang2024,Wang2025c,Sohal2025,Fan2024,Li2025c,Lee2025d,Lyons2024,Niwa2025,Chen2024,Lee2025,Li2021a,Chubb2021c}. This line of work has motivated the search for diagnostic tools that capture intrinsic structural properties of the decohered state itself \cite{Sang2024,Wang2025c,Sohal2025,Fan2024,Li2025c,Lee2025d}. According to recent studies, the mixed state corresponding to a noisy toric code shows a rich structure when one considers postselected quantum error correction \cite{English2025a}. In particular, while the non-postselected case is mapped to the RBIM, the fully postselected sector, corresponding to a zero-syndrome measurement outcome, is mapped to a clean Ising model \cite{English2025b}. This suggests that mappings from classical spin systems to quantum error-correcting codes play a key role in studying noisy quantum codes.

Besides the above classical-quantum mappings, there is also an interesting correspondence in which the partition function of a classical statistical model can be written as the inner product of a Calderbank-Shor-Steane (CSS) code state and a product state \cite{DenNest2007b,Zarei2018a}. This quantum formalism was first introduced in the context of measurement-based quantum computation in order to find complete models in statistical mechanics \cite{Nest2008a,Bravyi2007}. It has also been shown that this formalism can be used to identify phase transitions in noisy topological CSS codes \cite{Zarei2020d,Zarei2019e,Zarei2019a}. In this regard, the mapping of partition functions provides a mathematical framework for studying phase transitions in noisy topological quantum codes. In particular, it is an important task to find a quantum formalism for different well-known classical systems and to consider the corresponding properties of the associated noisy topological quantum codes.

One of the most well-known classical systems is the $XY$ model, which shows a Kosterlitz-Thouless (KT) phase transition. Since the $XY$ model does not have any ferromagnetic order at finite temperature, unlike the Ising model, it is of interest to consider which quantum code is mapped to the partition function of the $XY$ model. Given the continuous nature of the $XY$ model with $U(1)$ symmetry, it is natural to expect a connection to quantum codes based on continuous variables such as rotor- and oscillator-based quantum error-correcting codes which have been proposed as promising platforms \cite{Gottesman2001,Albert2017}. The toric-rotor code is an important example that illustrates significant challenges regarding the error threshold due to the continuity of noise \cite{Vuillot2024,Xu2024b}. In this paper, we develop a quantum formalism for the partition function of the $XY$ model and show that it is mapped to the inner product of a toric-rotor code state and a product state. We then use this formalism to consider how well-known facts about the KT transition in the $XY$ model are reflected in the physical properties of a noisy toric-rotor code.

We consider a logical state of toric-rotor code in the presence of phase-shift noise described by a von Mises probability distribution \cite{Vuillot2024}. We then show that the fidelity between the noisy mixed state and the initial logical state is proportional to the partition function of the $XY$ model, in the sense that the temperature on the classical side is mapped to the width of the noise on the quantum side. Accordingly, corresponding to the Kosterlitz-Thouless (KT) phase transition in the $XY$ model, we identify a resilience phase transition in the zero-syndrome postselected sector of the noisy toric-rotor code: the density matrix corresponding to the logical subspace, conditioned on a zero-syndrome measurement outcome, shows a transition from a partially coherent phase to a fully decoherent phase.

In order to characterize the nature of the different phases of the noisy toric-rotor code, we identify a topological order parameter corresponding to the well-known order parameters of the $XY$ model. In particular, we consider the spin stiffness of the $XY$ model as the order parameter, which is nonzero in the KT phase and drops discontinuously to zero in the paramagnetic phase. We then use the quantum formalism to map the spin stiffness in the $XY$ model to a topological resilience order parameter, $0\leq \lambda \leq 1$, in the noisy toric-rotor code. Regarding the above mapping, we conclude that below a critical noise width $\sigma_c \approx 0.89$ the resilience order parameter satisfies  $0< \lambda < 1$ which indicates partial coherence in the zero-syndrome subspace of the code, and drops discontinuously to zero at $\sigma_c$.  We finally give an argument that our approach is useful for considering postselected quantum error correction in the toric-rotor code in higher dimensions.

In Sec.~\ref{sec1}, we give a brief introduction to the toric-rotor code. In Sec.~\ref{sec2}, we introduce our quantum formalism for the partition function of the $XY$ model. In Sec.~\ref{sec3}, we define our noisy model for the toric-rotor code and demonstrate the correspondence between the resilience phase transition in the toric-rotor code and the KT phase transition in the $XY$ model. Finally, in Sec.~\ref{sec4}, we map the spin stiffness in the $XY$ model to a resilience order parameter that characterizes the different phases of the model, and in Sec.\ref{sec5}, discuss the connection between our results and postselected quantum error correction.

\section{Toric-rotor code}
\label{sec1}
In this section, we provide a brief introduction to the toric-rotor code. While the toric code is traditionally defined as a many-body system of qubits, the toric-rotor code is obtained by replacing qubits with quantum rotors. The simplest model of a quantum rotor, also known as a U(1) rotor, is a particle constrained to move along a circular path, such that its position is described by a phase. Another example is a superconducting circuit, in which the phase of the wave function plays a role analogous to the angular position of a particle on a circle \cite{Martinis2004}. The Hilbert space of the rotor is defined by its canonical variables, including the position variable \(\hat{\theta}\) (or phase), which is continuous and compact, and its conjugate angular-momentum variable \(\hat{\ell}\), which is discrete and unbounded \cite{Albert2017}. The commutation relation associated with these variables is \([\hat{\theta},\hat{\ell}]=i\).
This space is described by two orthogonal bases: an angular-position basis \(\{\ket{\theta} \mid \theta \in [0,2\pi) \}\) and an angular-momentum basis \(\{ \ket{\ell} \mid \ell \in \mathbb{Z} \}\).
The connection between these two bases is established by the following relations:
\begin{equation}
	\ket{\theta} = \frac{1}{\sqrt{2\pi}} \sum_{\ell \in \mathbb{Z}} e^{-i \ell \theta} \ket{\ell}\,,  \quad \ket{\ell}=\frac{1}{\sqrt{2\pi}} \int_{0}^{2\pi} d\theta
	e^{i \ell \theta} \,\ket{\theta}
\end{equation}
which are similar to Fourier transformations between position and momentum in harmonic oscillators. Similar to the harmonic oscillator, there are also displacement operators which lead to a shift in angular-position or angular-momentum bases. 
In particular, the momentum-displacement operator \(X(m)=e^{i m \hat{\theta}}\) is labeled by an integer \( m\in \mathbb{Z}\), whereas the phase-displacement operator \(Z(\phi)=e^{i \phi \hat{\ell}}\) is characterized by a continuous angle \(\phi\in[0,2\pi]\). The action of these operators on the system’s bases is as follows:
\begin{equation}
	\begin{aligned}
		X(m)\ket{\theta} &= e^{i m \theta}\ket{\theta}, &\qquad X(m)\ket{\ell} &= \ket{\ell+m} \quad \\[4pt]
		Z(\phi)\ket{\theta} &= \ket{\theta-\phi}, &\qquad Z(\phi)\ket{\ell} &= e^{i\phi \ell}\ket{\ell} \quad 
	\end{aligned}
\end{equation}
In this regard, \(X(m)\) and \(Z(\phi)\) have roles similar to Pauli operators and we call them generalized Pauli operators. The commutation relation between these operators, which is very canonical in rotor-based error-correction codes and governs the control of logical operators in this system, is given by the following expression:
\begin{equation}
	X(m)Z(\phi)=e^{-im\phi}Z(\phi)X(m)
\end{equation}
This relation shows that the operators \(X(m)\) and \(Z(\phi)\) commute only when \(m\phi=0 \,(\bmod\, 2\pi)\).
\begin{figure}[h!]  
	\centering
	\includegraphics[width=8.9cm,height=5.3cm,angle=0]{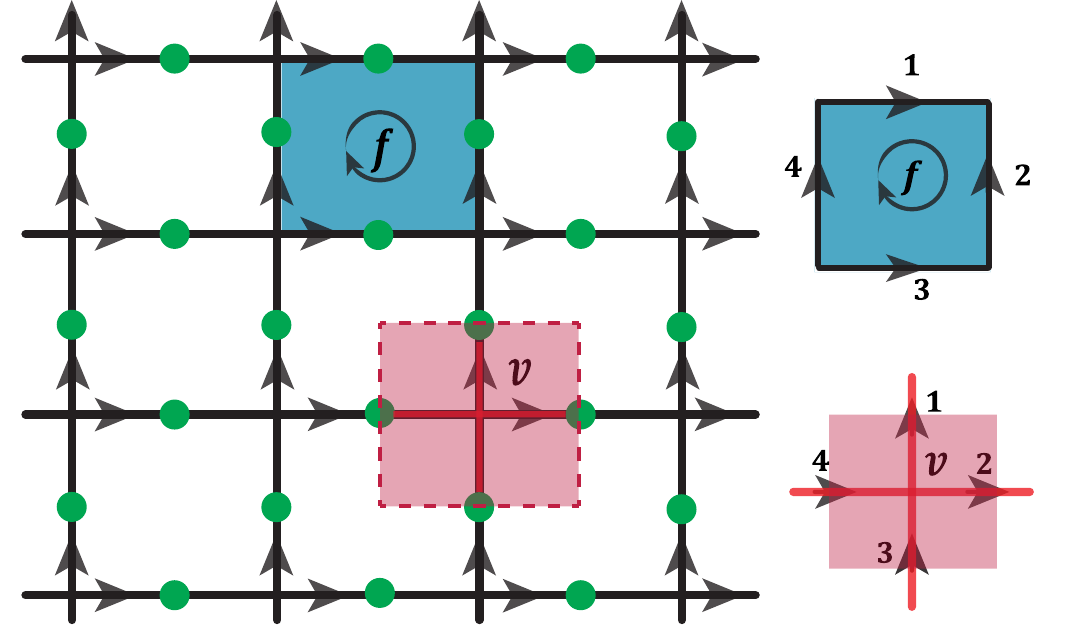} 
	\caption{Two-dimensional square lattice of the toric-rotor code. Quantum rotors (green circles) reside on the edges of the lattice, with a fixed orientation assigned to each edge. A blue square represents a face stabilizer, constructed as the product of the corresponding operators around the face; for example,
		$B_f(m)=X_1(m)X_2(-m)X_3(-m)X_4(m)$.
		A red square represents a vertex stabilizer, constructed as the product of the corresponding operators on the edges incident to the vertex; for example,
		$A_v(\phi)=Z_1(\phi)Z_2(\phi)Z_3(-\phi)Z_4(-\phi)$.}
	\label{TR}	
\end{figure}
Now, we are ready to introduce a toric-rotor code which is typically defined on a two-dimensional square lattice with periodic boundary conditions (i.e., a torus). In this model, each edge of the lattice hosts a quantum rotor instead of a qubit, and the code structure is determined by the stabilizer group, whose generators are the face and vertex stabilizers. In order to define stabilizers, we assign an orientation to each edge of the lattice as shown in Fig.~\ref{TR}. Then face stabilizers are constructed as the product of \(X(m)\) operators around each face of the lattice:
\begin{equation}
	B_f(m)=\prod_{e\in\partial{f}}X_e(m)^{\varepsilon_{e,f}}=e^{\,i m \sum_{e\in\partial f}\varepsilon_{e,f}\,\hat{\theta}_e\,}
\end{equation}
where \(\varepsilon_{e,f}=\pm1\) specifies the orientation of an edge, such that \(\varepsilon_{e,f}\) is set to $+1$ ($-1$) if the orientation of the corresponding edge matches clockwise (anticlockwise) circulation around the associated face; see Fig.~\ref{TR}. Vertex stabilizers are constructed as the product of \(Z(\phi)\) operators incident on each vertex:
\begin{equation}
	A_v(\phi)=\prod_{e\in v}Z_e(\phi)^{\varepsilon_{e,v}}=e^{\,i \phi \sum_{e\in v}\varepsilon_{e,v}\,\hat{\ell}_e\,}
\end{equation}
where \(\varepsilon_{e,v}=\pm1\) is set to $-1$ ($+1$) if the corresponding edge is incoming to (outgoing from) the associated vertex; see Fig.~\ref{TR}. The specific definition of stabilizers according to orientation of the lattice edges ensures that the vertex and face stabilizer operators commute, that is, \([A_v(\phi),B_f(m)]=0\). In particular, it is simple to check that for any choice of edge orientations, the commutation relation holds. However, in this work, we restrict our attention to the specific orientation shown in Fig.~\ref{TR}.

Due to the torus topology, the subspace of the code includes two logical rotors. In order to characterize logical states of the code, note that the logical subspace consists of all states that are stabilized by every vertex \(A_v(\phi)\) and face stabilizer \(B_f(m)\), with eigenvalue \(+1\).
Starting from the configuration in which all rotors have angular momentum \(\ell=0\), one of the logical states can be expressed in the following unnormalized form:
\begin{equation}\label{ketG}
	\begin{aligned}
		\lvert G \rangle
		&= \prod_{f} \left( \sum_{n_f \in\mathbb{Z}} B_f(n_f) \right) \lvert \ell = 0 \rangle^{\otimes M} \\
		&= \prod_{f} \left( \sum_{n_f \in\mathbb{Z}} e^{i n_f \sum_{e \in \partial f} \varepsilon_{e,f}\,\hat{\theta}_e} \right)
		\lvert \ell = 0 \rangle^{\otimes M}
	\end{aligned}
\end{equation}
Here $M$ refers to the number of rotors, $\prod_{f}$ refers to a product corresponding to all independent faces of the lattice, and $n_f \in\mathbb{Z}$ is an integer corresponding to each face $f$. The above state can also be defined as the limit of a normalized state. As discussed in \cite{Royer2020,Xu2024b}, one can use a standard regularization procedure by applying a Gaussian envelope to the ideal state:
\begin{equation}\label{ketphys}
	\ket{G_{\mathrm{phys}}}
	=
	\mathcal{N}_{\Delta}
	\prod_{f}
	\left(
	\sum_{n_{f}\in\mathbb{Z}}
	e^{-\frac{\Delta}{2} n_{f}^{2}}\, B_f(n_{f})
	\right)
	\ket{\ell=0}^{\otimes M},
\end{equation}
where $\Delta > 0$ is the regularization parameter and $\mathcal{N}_{\Delta}$ is a normalization constant proportional to a Jacobi theta function. In the limit $\Delta \to 0$, we recover the ideal state given in Eq.~(\ref{ketG}).

\begin{figure}[h!]  
	\centering
	\includegraphics[width=9.2cm,height=7.3cm,angle=0]{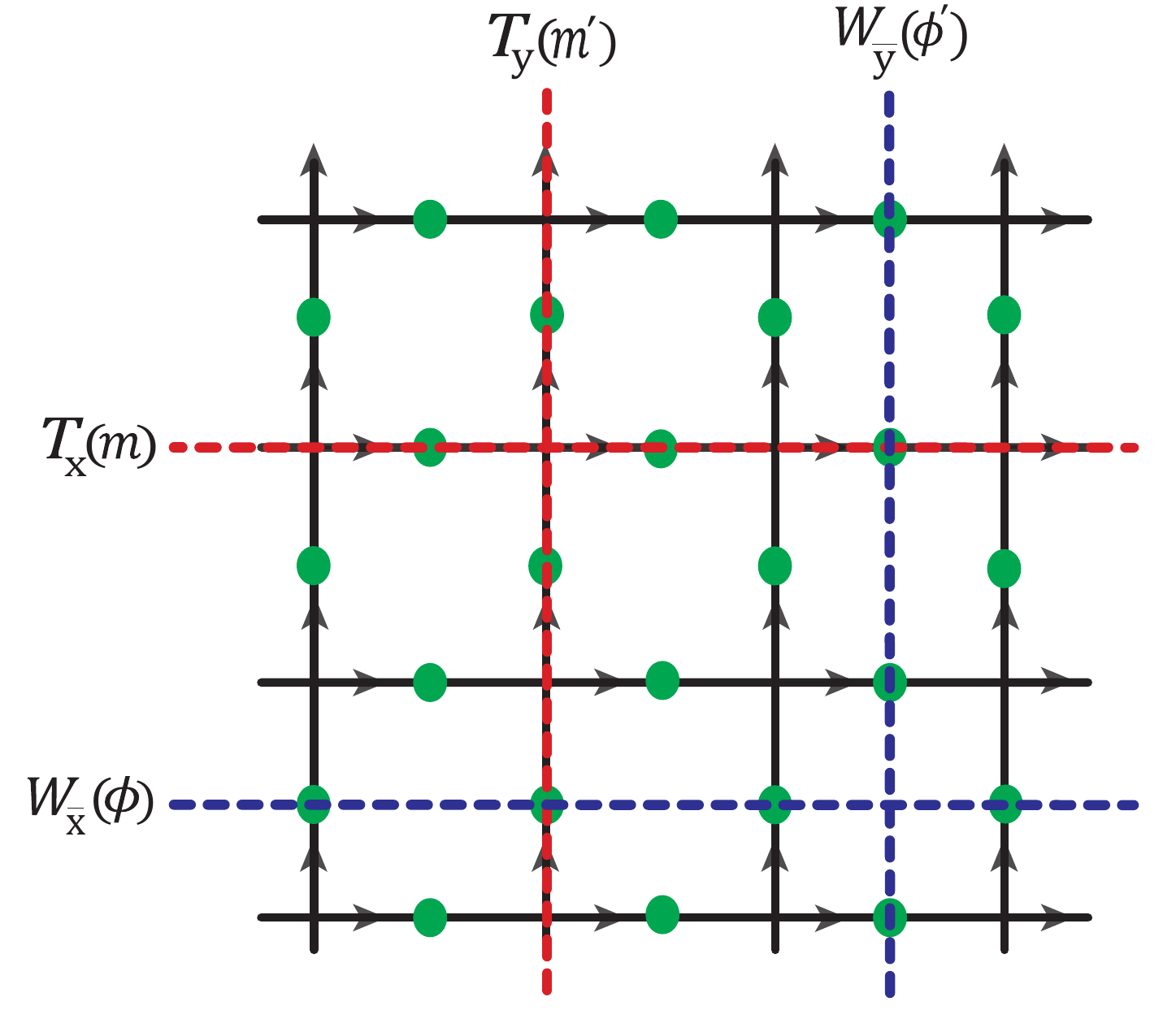} 
	\caption{Non-local operators in the toric-rotor code. The red lines denote the non-contractible loops $T_x(m)$ and $T_y(m')$ along the horizontal and vertical lattice directions, formed by products of the corresponding operators. The blue lines denote the operators $W_{\bar{x}}(\phi)$ and $W_{\bar{y}}(\phi')$, defined along horizontal and vertical paths of the dual lattice and intersecting the non-contractible loops.}
	\label{P2}
\end{figure}

Now let us go back to the ideal state. Owing to the topology of the torus, there exist two operators corresponding to non-contractible loops around the torus that commute with all stabilizers but are not members of the stabilizer group, i.e., they cannot be generated by any product of stabilizers.  As shown in Fig.~\ref{P2}, we denote these non-trivial loop operators by \(T_x(m)\) and \(T_y(m')\):
\begin{equation}
	\begin{aligned}
		T_x(m)
		&=\prod_{e\in \mathcal{C}_x}X_e(m)=e^{\,i m \sum_{e\in\mathcal{C}_x}\hat{\theta}_e\,} \\
		T_y(m')
		&=\prod_{e\in \mathcal{C}_y}X_e(m')=e^{\,i m' \sum_{e\in\mathcal{C}_y}\hat{\theta}_e\,}
	\end{aligned}
\end{equation}
Here, \(\mathcal{C}_x\) and \(\mathcal{C}_y\) denote the non-contractible loops along the horizontal and vertical directions of the torus, respectively, and \(m\) and \(m'\) are integers.
Applying these operators to the reference logical state \(\ket{G}\) maps the system to another orthogonal sector within the code space.
The other logical states of the toric-rotor code can be generated from them as follows:
\begin{equation}
	|\Phi_{m,m'}\rangle = T_x(m)\,T_y(m')\,\ket{G}
\end{equation}
Unlike qubit codes, in which the dimension of the code subspace is finite, the number of logical states in this model is countably infinite, reflecting the compact nature of the position variable \(\hat{\theta}\) and the unbounded spectrum of the momentum variable \(\hat{\ell}\). Each pair \((m,m')\) specifies a distinct topological sector.
Since no local operator can distinguish these states from one another, one should define the dual non-local operators \(W_{\bar{x}}(\phi)\) and \(W_{\bar{y}}(\phi')\) on paths that intersect the non-contractible loops associated with the \(T\) operators, as shown in Fig.~\ref{P2}, as follows: 
\begin{equation}\label{w}
	\begin{aligned}
		W_{\bar{x}}(\phi)
		&= \prod_{e \in C_{\bar{x}}} Z_e(\phi)
		= e^{i \phi \sum_{e \in C_{\bar{x}}} \hat{\ell}_e} , \\
		W_{\bar{y}}(\phi')
		&= \prod_{e \in C_{\bar{y}}} Z_e(\phi')
		= e^{i \phi' \sum_{e \in C_{\bar{y}}} \hat{\ell}_e} 
	\end{aligned}
\end{equation}
where \(C_{\bar{x}}\) and \(C_{\bar{y}}\) denote non-contractible loops on the dual lattice, acting only on the edges intersected by the corresponding primal loops.
In the basis \(\ket{\Phi_{m,m'}}\), the \(T\) operators act as shift operators between different topological sectors, whereas the action of the \(W\) operators on this state yields a phase proportional to the corresponding integer label. Their action is given by: 
\begin{equation}
	\begin{aligned}
		T_x(k)\ket{\Phi_{m,m'}}&=\ket{\Phi_{m+k,m'}} \\
		T_y(k')\ket{\Phi_{m,m'}}&=\ket{\Phi_{m,m'+k'}} \\
		W_{\bar{x}}(\varphi)\ket{\Phi_{m,m'}}&=e^{im'\varphi}\ket{\Phi_{m,m'}} \\
		W_{\bar{y}}(\varphi')\ket{\Phi_{m,m'}}&=e^{im\varphi'}\ket{\Phi_{m,m'}}
	\end{aligned}
\end{equation}
We also note that the code space can be expressed in terms of the basis constructed from the \(W\) operators. Since in the toric-rotor code the Hilbert space has a composite structure, one may employ a continuous phase basis ($\phi$) to store information rather than the discrete momentum basis \((m)\). By summing over all integers \(m\) and \(m'\) with the weights \(e^{im\phi}\) and \(e^{im'\phi'}\), we obtain a new basis \(\ket{\Psi_{\phi,\phi'}}\), in which the roles of the \(T\) and \(W\) operators are interchanged. This basis is given by:
\begin{equation}\label{gs}
	\ket{\Psi_{\phi,\phi'}}=\sum_m e^{im\phi}T_x(m)\sum_{m'} e^{im'\phi'}T_y(m')\ket{G}
\end{equation}
In this basis, the \(T\) operators contribute only a phase, whereas the \(W\) operators act as shift operators:
\begin{equation}\label{wket}
	\begin{aligned}
		T_x(k)\ket{\Psi_{\phi,\phi'}}&=e^{-ik\phi}\ket{\Psi_{\phi,\phi'}} \\
		T_y(k')\ket{\Psi_{\phi,\phi'}}&=e^{-ik'\phi'}\ket{\Psi_{\phi,\phi'}} \\ 
		W_{\bar{x}}(\varphi)\ket{\Psi_{\phi,\phi'}}&=\ket{\Psi_{\phi,\phi'+\varphi}} \\
		W_{\bar{y}}(\varphi')\ket{\Psi_{\phi,\phi'}}&=\ket{\Psi_{\phi+\varphi',\phi'}}		
	\end{aligned}
\end{equation}

In other words, the above non-local operators play the role of logical generalized Pauli gates in the logical subspace. We emphasize that the topological sectors are protected by topology, since it is highly unlikely for a local perturbation to generate such operators. However, because of the continuity of the phase variable, even an operator with a very small weight can act as a logical gate, thereby posing significant challenges for error correction in such codes.

\section{Mapping to classical XY model}
\label{sec2}
\begin{figure}[h!]  
	\centering
	\includegraphics[width=7.0cm,height=6.4cm,angle=0]{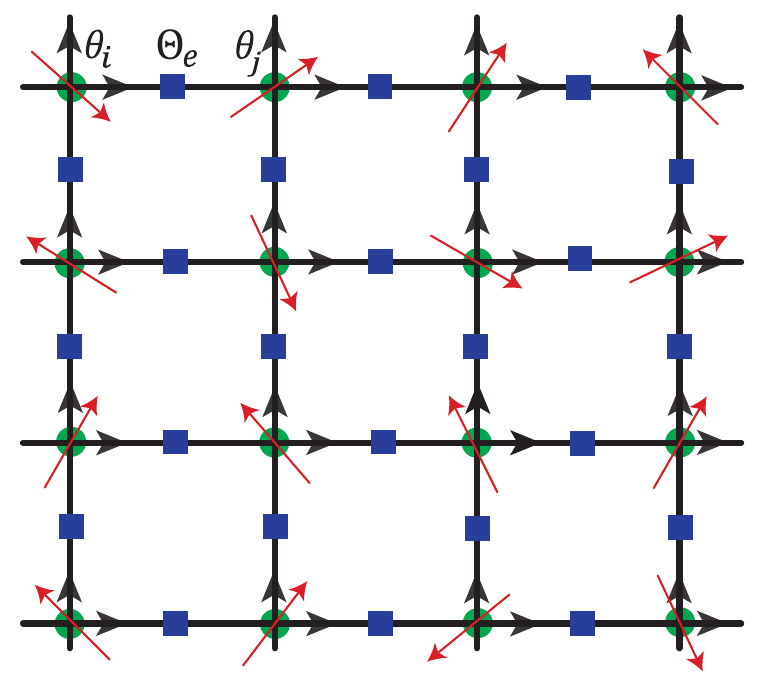} 
	\caption{Square-lattice representation of the two-dimensional $XY$ model. An orientation is assigned to each lattice edge, and the partition function is rewritten by replacing vertex variables $\theta$ with edge variables $\Theta_e$. Each edge variable $\Theta_e$ (blue squares) is associated with an oriented edge $e$ connecting vertices $i$ and $j$.}
	\label{XY}
\end{figure}

\begin{figure}[h!]  
	\centering
	\includegraphics[width=7.4cm,height=6.6cm,angle=0]{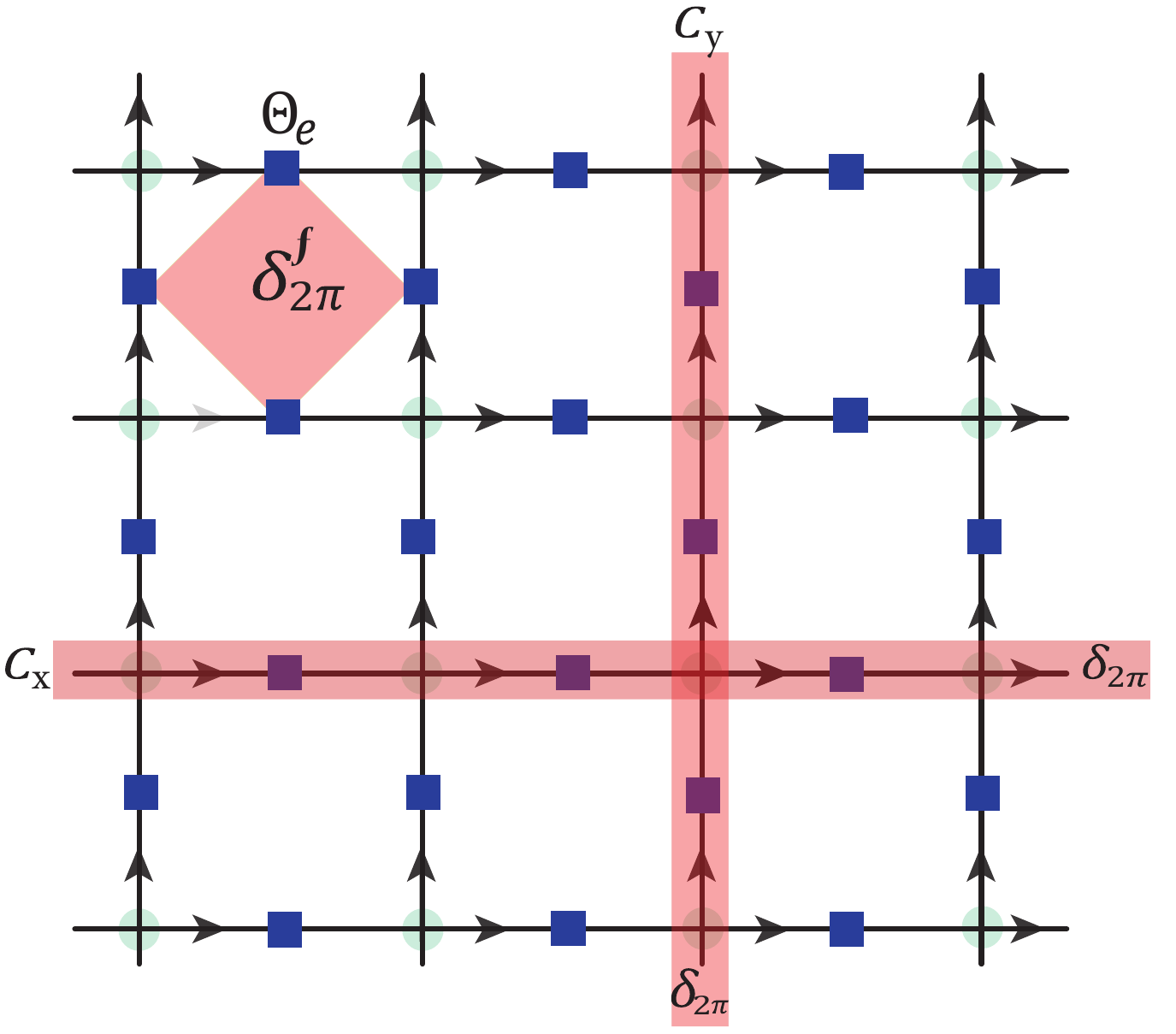} 
	\caption{Geometric representation of the constraints on the edge variables in the rewritten partition function [Eq.~(\ref{partitionfunction})]. The symbol $\overset{f}{\delta_{2\pi}}$ (diamond-shaped region) denotes the local plaquette constraint, enforcing that the oriented sum of edge variables around any closed loop equals an integer multiple of $2\pi$. The horizontal and vertical colored bands indicate nonlocal constraints imposed by periodic boundary conditions.}
	\label{P4}
\end{figure}

In order to investigate a quantum-classical correspondence, we introduce a quantum formalism for the partition function of classical $XY$ model.
The Hamiltonian of this model is defined as follows:
\begin{equation}\label{HXY}
	H = -J \sum_{\langle ij\rangle} \cos(\theta_i - \theta_j)
\end{equation}
Here $\theta_i$ denotes the spin angle at vertex \(i\), as shown in Fig.~\ref{XY}, and the Hamiltonian involves a sum over nearest-neighbor pairs on the square lattice with periodic boundary conditions. The partition function of the model is given by
\begin{equation}
	\mathcal{Z} =  \int_{0}^{2\pi} \prod_{i=1}^N \frac{d\theta_i}{2\pi}\,
	\exp\!\Big( \beta \sum_{\langle ij\rangle} \cos(\theta_i - \theta_j)\Big)
\end{equation}
Here, $N$ refers to number of spins and $\beta$ denotes the inverse temperature $1/T$, where we set the Boltzmann constant $k_B$ to $1$. We assign an orientation to each edge of the lattice and, for each directed edge $e$, define a new variable \(\Theta_e \equiv \theta_i - \theta_j\), where $i$ and $j$ denote the vertices corresponding to the starting and ending points of the edge. We then associate these variables with the edges of the lattice and refer to them as edge variables. Next, we express the partition function in terms of these edge variables instead of the original vertex variables.
However, the edge variables \(\Theta_e\) are not independent and must satisfy several geometric constraints. These constraints are incorporated into the partition function through delta functions. The partition function can then be rewritten in terms of these new variables as follows:

\begin{equation}\label{partitionfunction}
	\mathcal{Z} =  \int_{-\pi}^{\pi} \prod_{e}^M \frac{d\Theta_e}{2\pi}\,
	e^{\beta  \sum_{e} \cos(\Theta_e)}
	\Big[\prod_{f}^{N-1} \overset{f}{\delta_{2\pi}}\Big]\Big[\overset{C_x}{\delta_{2\pi}}\Big]\Big[\overset{C_y}{\delta_{2\pi}}\Big]
\end{equation}
where \( \overset{f}{\delta_{2\pi}}= \delta_{2\pi}\!\Big(\sum_{e\in\partial f} \varepsilon_{e,f}\,\Theta_e\Big)\) is a periodic delta function which imposes the constraint that the sum of the edge variables around each face of the lattice (i.e., each closed loop) is equal to \(2\pi n\) with \(n \in \mathbb{Z}\), see Fig.~\ref{P4}. The factor \(\varepsilon_{e,f}=\pm1\) specifies the orientation of each edge in the traversal of the loop. Similarly, \(\overset{C_x}{\delta_{2\pi}}=\delta_{2\pi}\!\Big(\sum_{e\in C_x}\Theta_e\Big)\) and \(\overset{C_y}{\delta_{2\pi}}=\delta_{2\pi}\!\Big(\sum_{e\in C_y}\Theta_e\Big)\) impose the constraints associated with periodic boundary conditions. These conditions require that the sum of the edge variables along each non-contractible loop of the lattice be an integer multiple of \(2\pi\).
In the next step, we notice that the periodic delta function can be written as a discrete Fourier series,
\begin{align}
	\delta_{2\pi}(x) &= \sum_{m\in\mathbb{Z}} \delta(x - 2\pi m)
	= \frac{1}{2\pi}\sum_{q\in\mathbb{Z}} e^{i q x}
\end{align}
Using this representation, the partition function can be written in the form
\begin{equation}\label{eq:partition}
	\mathcal{Z} = \int_{-\pi}^{\pi} \prod_{e=1}^{M} \frac{d\Theta_e}{2\pi}\,
	\exp\!\Big(\beta \sum_{e=1}^{M}\cos(\Theta_e)\Big)\mathcal{W}_b\mathcal{W}_l
\end{equation}
where \begin{equation}
	\mathcal{W}_b= \prod_{f=1}^{N-1}\left(\frac{1}{2\pi}\sum_{n_f\in\mathbb{Z}}
	e^{i n_f \sum_{e\in\partial f} \varepsilon_{e,f}\,\Theta_e} \right)
\end{equation}
where we have replaced the delta functions corresponding to each face with a Fourier series, with $n_f$ denoting the variable associated with a face $f$, and
\begin{equation}
	\mathcal{W}_l= \frac{1}{4\pi^2}\left(\sum_{m\in\mathbb{Z}}
	e^{i m \sum_{e\in C_x} \Theta_e} \right)\left(\sum_{m'\in\mathbb{Z}}
	e^{i m' \sum_{e\in C_y}\Theta_e} \right)
\end{equation}
has been introduced in place of the delta functions corresponding to the non-trivial loops.

Now we introduce a quantum formalism for the above relation. To this end, we first use a simple lemma that the integral of an arbitrary function $f(\Theta)$ can be written in the following form:
\begin{equation}\label{eqf}
	\big\langle \ell=0 \big|\, f(\hat\Theta)\, \big| \ell=0 \big\rangle
	= \int_{-\pi}^{\pi} \frac{d\Theta}{2\pi}\,f(\Theta)
\end{equation}
where \(\ket{\ell=0}=
\frac{1}{\sqrt{2\pi}}\int_{-\pi}^{\pi} d\Theta \ket{\Theta}\) denotes the angular-momentum eigenstate with eigenvalue \(\ell=0\), and where \(\ket{\Theta}\) are the eigenstates of the angular operator \(\hat{\Theta}\) in the interval \([-\pi,\pi]\) and satisfy the relation \(\bra{\Theta'}\ket{\Theta}=\delta(\Theta'-\Theta)\).
This lemma is easily proved by applying the operator \(f(\hat\Theta)\) to the state \(\ket{\ell=0}\) and using the fact that \(f(\hat\Theta)\ket{\Theta}=f(\Theta)\ket{\Theta}\). It can also be extended to a function of several variables $f(\{\Theta_e\})$ in the following form:
\begin{equation}\label{quantumformalism}
	\int_{-\pi}^{\pi}\prod_{e=1}^{M}\frac{d\Theta_{e}}{2\pi} f(\{\Theta_{e}\})
	=  ~^{M\otimes}\langle \ell=0\big| f(\{\hat\Theta_{e}\}) \big|\ell=0\big\rangle^{\otimes M}.
\end{equation}
We now use the above formalism for the partition function in Eq.~(\ref{eq:partition}). In particular, notice that the terms \(\mathcal{W}_b\) and \(\mathcal{W}_l\) are replaced by the following quantum operators:
\begin{equation}
	\begin{aligned}
		\hat{\mathcal{W}}_b &= \prod_{f=1}^{N-1}\left(\frac{1}{2\pi}\sum_{n_f\in\mathbb{Z}}
		e^{i n_f \sum_{e\in\partial f} \varepsilon_{e,f}\,\hat{\Theta}_e} \right) \\
		\hat{\mathcal{W}}_l &= \frac{1}{4\pi^2}\left(\sum_{m\in\mathbb{Z}}
		e^{i m \sum_{e\in C_x} \hat{\Theta}_e} \right)\left(\sum_{m'\in\mathbb{Z}}
		e^{i m' \sum_{e\in C_y}\hat{\Theta}_e} \right)
	\end{aligned}
\end{equation}
and accordingly, the partition function is written in the form:
\begin{equation}\label{quantumformalism1}
	\mathcal{Z} = \langle \ell=0|
	e^{\,\beta\sum_{e}\cos\hat\Theta_e}\,
	\Bigg(\hat{\mathcal{W}}_l \hat{\mathcal{W}}_b
	\Bigg)
	\ket{\ell=0}^{\otimes M}
\end{equation}
Then we notice that $\hat{\mathcal{W}}_b \ket{\ell=0}^{\otimes M}$ is the same as the toric-rotor code state $|G\rangle$ in Eq.~(\ref{ketG}) up to a multiplication factor $1/(2\pi)^{N-1}$. On the other hand, $\hat{\mathcal{W}}_l$ can also be written in terms of logical operators $T_x (m)$ and $T_y (m')$ in the toric-rotor code in the form of $\hat{\mathcal{W}}_l =1/4\pi^2 \sum_{m,m'}T_x (m) T_y (m')$. In this regard, by comparing with Eq.~(\ref{gs}), it is simply concluded that $\hat{\mathcal{W}}_l |G\rangle$, up to a multiplication factor, is equal to the logical state $|\Psi_{0,0}\rangle$ in the toric-rotor code. Finally, the partition function of the $XY$ model is related to the toric-rotor code state as follows
\begin{equation}\label{eq1}
	\mathcal{Z} = \frac{1}{(2\pi)^{N+1}}\bra{\alpha}\ket{\Psi_{0,0}}
\end{equation}
where \(\ket{\alpha}=\prod_{e} e^{\beta  \cos \hat{\Theta}_e} \ket{\ell=0}\) is a product state. 
In this formulation, the partition function of classical $XY$ model is expressed as the inner product of a product state and a logical state of the toric-rotor code. We note that the classical $XY$ model has been extensively investigated in the context of topological order in classical systems, where it exhibits a KT phase transition at a critical temperature. The main question, then, is how to interpret the right-hand side of Eq.~(\ref{eq1}) when a logical state of the toric-rotor code appears. To this end, we write \(\ket{\alpha}\) as
\begin{equation}
	\ket{\alpha}= \frac{1}{\sqrt{2\pi}} \int_{-\pi}^{\pi} d\Theta
	e^{\beta \cos \Theta}\ket{\Theta}
\end{equation}
If we express \(\ket{\Theta}\) as \(\ket{\Theta}=e^{i\Theta\hat{\ell}}\ket{\Theta=0}\), then the state \(\ket{\alpha}\) becomes
\begin{equation}
	\ket{\alpha}= \frac{1}{\sqrt{2\pi}} \bigg[\int_{0}^{2\pi} d\Theta
	e^{\beta  \cos \Theta}e^{i\Theta\hat{\ell}}\bigg]\ket{\Theta=0}
\end{equation}
According to this representation, the effect of state \(\ket{\alpha}\) on the logical ground state \(\ket{\Psi_{0,0}}\) can be interpreted as applying noise $e^{i\Theta\hat{\ell}}$, characterized by a specific probability distribution proportional to $e^{\beta  \cos \Theta}$. We investigate this interpretation in the next section by considering the toric-rotor code in the presence of noise. 
\section{Phase transition in noisy toric-rotor code}
\label{sec3}
In this section, we consider toric-rotor code in the presence of a particular type of noise. We consider phase-shift noise which is described by applying generalized Pauli operator \(Z\) which shifts the phase of a quantum state $\ket{\phi}$ by an angle \(\Theta\):
\begin{equation}
	Z(\Theta)=e^{i \Theta \hat{\ell}}
\end{equation}
Unlike qubit systems, where errors are typically discrete, the phase variable \(\Theta\) in rotor systems is continuous, so that any nonzero value of \(\Theta\) constitutes an error. However, different phase shifts do not occur with equal probability. We assume that each rotor independently experiences a \(Z\)-type error distributed according to a probability density \(P(\Theta)\).
Since small phase errors are much more likely than large phase errors (i.e., those close to \(\pi\)), the von Mises probability distribution \cite{Vuillot2024} provides a physically appropriate model for continuous phase noise and takes the form:
\begin{equation}
	P(\Theta) = \frac{e^{\frac{ \cos{\Theta}}{\sigma}}}{2\pi I_0(1/\sigma)}
\end{equation}
where \(I_0(1/\sigma)\) denotes the modified Bessel function of the first kind of order zero, which appears in the normalization factor. This distribution is well suited for modeling phase errors in rotor-based codes because it respects the periodicity of the phase variable \(\Theta\) and reduces to a Gaussian distribution in the limit of small $\sigma$ and correspondingly small $\Theta$.

We now assume that the initial state of the system is \(\rho_0 = \ket{\Psi_{0,0}}\bra{\Psi_{0,0}}\). Applying noise to the system is described by a quantum channel in the following Kraus representation:
\begin{equation}\label{rho0}
	\mathcal{E}(\rho_0) =
	\int \prod_e d\Theta_e \, P_e(\Theta_e) \,\prod_e Z_e(\Theta_e) \rho_0
	\prod_eZ_e^{\dagger}(\Theta_e)
\end{equation}
An important quantity is the probability that the system remains in the state \(\ket{\Psi_{0,0}}\) after applying the noise. This probability can be obtained by computing the fidelity between the noisy state and the original state \(\ket{\Psi_{0,0}}\), which is defined as
\begin{equation}\label{eq00}
	\mathcal{F}=\bra{\Psi_{0,0}}\mathcal{E}(\rho_0)\ket{\Psi_{0,0}}
\end{equation}
This expression is nonzero only when the phase variables \(\{\Theta_e\}\) satisfy specific constraints such that the operator \(\prod_e Z_e(\Theta_e)\) belongs to the stabilizer group of the state $|\Psi_{0,0}\rangle $. To this end, \(\prod_e Z_e(\Theta_e)\) must commute with the stabilizer operators \(B_f(m)\) as well as the logical operators \(T_x(m)\) and \(T_y(m')\). The requirement that \(\prod_e Z_e(\Theta_e)\) commutes with each of these operators imposes corresponding constraints, which are enforced through periodic delta functions. 
As a result, the fidelity can be written as
\begin{equation}
	\mathcal{F}= \frac{1}{[2\pi I_0(1/\sigma)]^M}\int \prod_e d\Theta_e e^{\frac{ \sum_{e} \cos{\Theta_e}}{\sigma}} \Big[\prod_{f}^{N-1} \overset{f}{\delta_{2\pi}}\Big]\Big[\overset{C_x}{\delta_{2\pi}}\Big]\Big[\overset{C_y}{\delta_{2\pi}}\Big]
\end{equation}
By comparing with Eq.~(\ref{partitionfunction}) and substituting $1/\sigma =\beta$, this expression is proportional to the partition function of the two-dimensional $XY$ model. Therefore, we have:
\begin{equation}
	\mathcal{F}= \frac{1}{[I_0(1/\sigma)]^M}\mathcal{Z}
\end{equation}
Because the partition function of the two-dimensional $XY$ model, and hence its free energy, exhibits an essential singularity at the KT transition temperature \(T_c\approx0.89\) \cite{Kosterlitz1973a,Gupta1992}, the corresponding quantum state fidelity is expected to display analogous non-analytic behavior at a critical width $\sigma_c$. The quantum model therefore undergoes a KT-type phase transition and we should identify a physical quantity that characterizes this transition. Guided by the established mapping, we consider the order parameter of the classical $XY$ model and then look for its quantum analogue. First, we notice that by the Mermin-Wagner theorem \cite{Mermin1966a}, systems with continuous symmetry and short-range interactions in \(d \leq 2\) lack long-range order at any finite temperature, implying vanishing magnetization for all \(T>0\). As a result, magnetization cannot distinguish the low-temperature phase with quasi-long-range order from the high-temperature disordered phase in the $XY$ model.

\begin{figure}[h!]  
	\centering
	\includegraphics[width=8.0cm,height=7.0cm,angle=0]{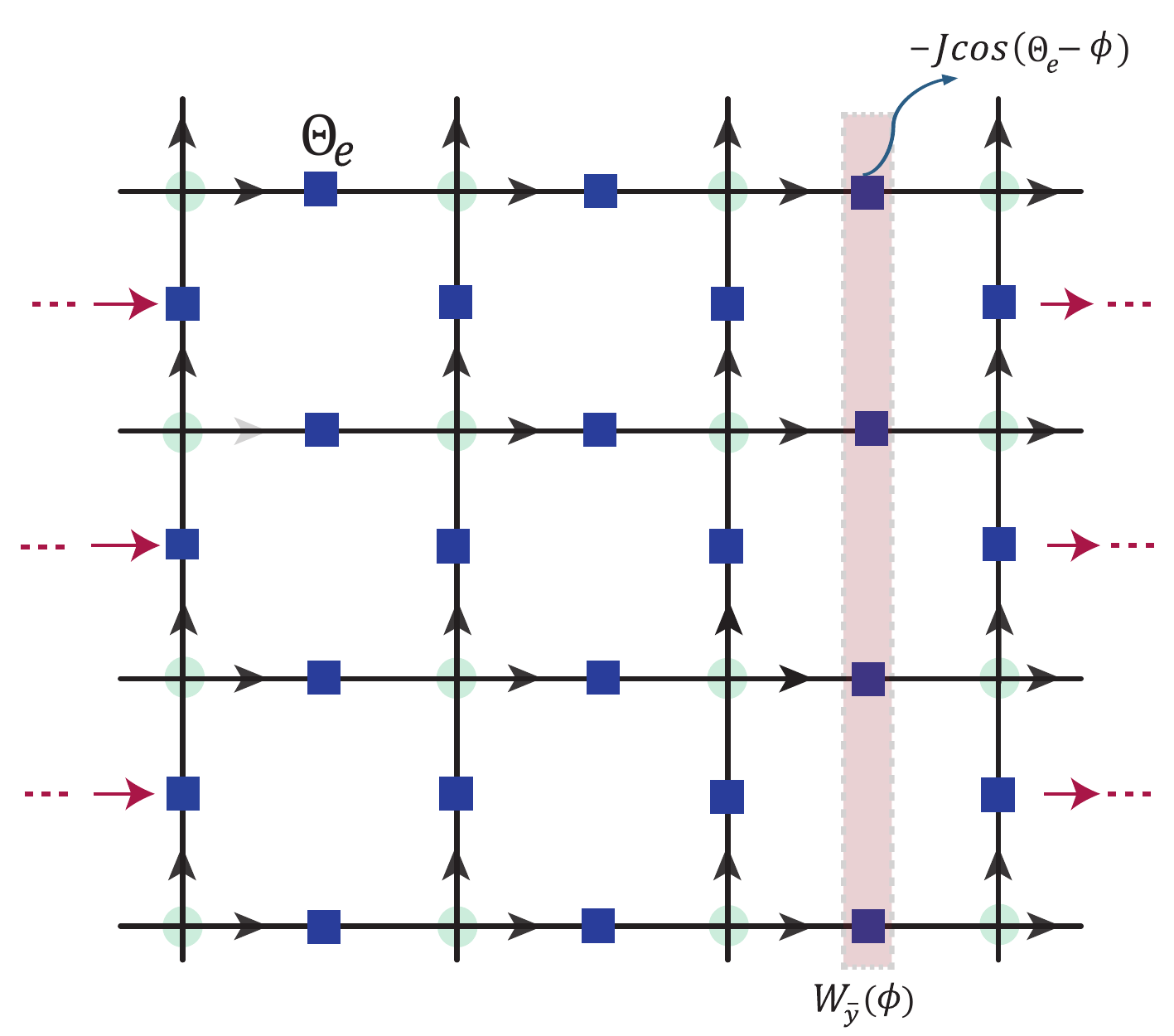} 
	\caption{Schematic illustration of twisted periodic boundary conditions for the $XY$ model and its interpretation in the toric-rotor code. The twist is introduced by imposing a phase shift $\phi$ on the edges crossing a specified boundary (indicated by the red shaded bar corresponding to a non-contractible loop $C_{\bar{y}}$). For these edges, which $\delta_e = 1$, the interaction term in the Hamiltonian changes to $-J \cos(\Theta_e - \phi)$ where $\Theta_e =\theta_i -\theta_j$. Using the quantum formalism, the twist corresponds to applying $W_{\bar{y}}(\phi)$ to the initial state $|\Psi_{0,0}\rangle$ in the presence of noise.}
	\label{twistw}
\end{figure}

The KT transition is instead characterized by the spin stiffness (helicity modulus), which quantifies the system's response to a twist in the boundary conditions \cite{Hsieh2013a}. To define the spin stiffness, one considers a Hamiltonian similar to Eq.~(\ref{HXY}) with a twist introduced by modifying the boundary couplings to \(-J\cos(\theta_i - \theta_j -\phi)\), as shown in Fig.~\ref{twistw}. Here, \(\phi\) is the twist angle, and the original model is recovered for \(\phi = 0\). The twisted Hamiltonian can be written as:
\begin{equation}\label{xys}
	H_{\phi} = -J \sum_{e} \cos(\Theta_e - \delta_e \phi),
\end{equation}
where $\Theta_e=\theta_i -\theta_j $, \(\delta_e = 1\) for edges \(e\) corresponding to the twisted boundary and \(\delta_e = 0\) otherwise. The partition function and free energy of the twisted model are functions of $\phi$, where $\mathcal{A}_\phi = -T\ln \mathcal{Z}_\phi$.

 In a macroscopic phase with quasi-long-range order, imposing a twist in the boundary conditions leads to an increase in the free energy of the system $\Delta \mathcal{A} = \mathcal{A}_\phi - \mathcal{A}_0$. For a given angle $\phi$, spin stiffness is defined in terms of this increase as follows: 
\begin{equation}\label{rhos1}
	\rho_s(\phi) = \frac{2\,\Delta \mathcal{A}}{\phi^2} L^{2-d}.
\end{equation}
where \(d\) is the spatial dimension of the system, $L$ is the linear system size and \(\rho_s\) denotes the spin stiffness \cite{Khairnar2025}. In the limit \(\phi \rightarrow 0\), this quantity can be expressed as
\begin{equation}\label{rhos2}
	\rho_s = \left( \frac{\partial^2 \mathcal{A}}{\partial \phi^2} \right)_{\phi=0} L^{2-d}
\end{equation}

\begin{figure}[h!]  
	\centering
	\includegraphics[width=7.5cm,height=5.3cm,angle=0]{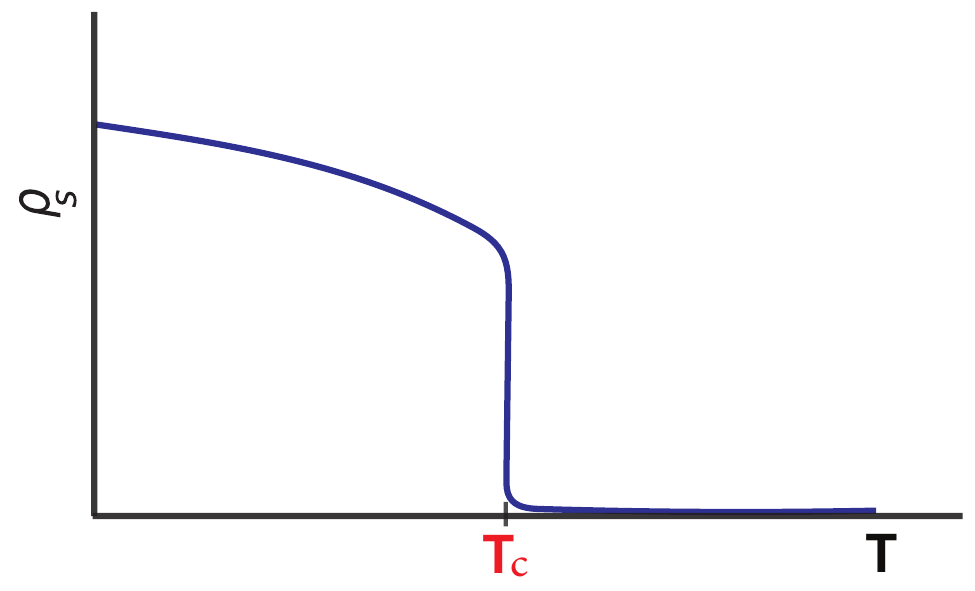} 
	\caption{Spin stiffness $\rho_s$ as a function of temperature $T$ for the two-dimensional $XY$ model \cite{Gerber2015}. At low temperatures ($T<T_c$), $\rho_s$ remains finite, reflecting the system's elastic response to a boundary twist. In the thermodynamic limit, $\rho_s$ exhibits a universal discontinuous jump from a finite value to zero at the critical temperature $T_c\approx0.89$, signaling the transition from the quasi-long-range ordered phase to the disordered phase. }
	\label{P5}
\end{figure}

In the disordered phase, the system does not resist boundary twists and the spin stiffness vanishes, \(\rho_s=0\), whereas in the ordered phase elastic spin correlations lead to a finite stiffness, \(\rho_s\neq0\). In two dimensions, it has been shown \cite{Minnhagen2003} that \(\rho_s\) exhibits a discontinuous jump from a finite value to zero at the transition temperature \(T_{c}\) (Fig.~\ref{P5}).

\section{Spin stiffness and a resilience order parameter}
\label{sec4}
Since spin stiffness is a suitable order parameter for $XY$ model, in this section we look for its quantum analogue by using the classical-to-quantum mapping. To this end, we focus on the partition function of the twisted model Eq.~(\ref{xys}) and find a quantum formalism. We write the partition function in the presence of a twist analogously to that of the untwisted case:
\begin{equation}\label{patw}
	\mathcal{Z}_\phi = \int_{-\pi}^{\pi} \prod_{e}^M \frac{d\Theta_e}{2\pi}\,
	e^{\beta  \sum_{e} \cos(\Theta_e-\delta_e\phi)}
	\Big[\prod_{f}^{N-1} \overset{f}{\delta_{2\pi}}\Big]\Big[\overset{C_x}{\delta_{2\pi}}\Big]\Big[\overset{C_y}{\delta_{2\pi}}\Big]
\end{equation}
Then, similar to Eq.~(\ref{quantumformalism1}) for the untwisted model, the partition function of the twisted model can be written in quantum formalism as follows:
\begin{equation}
	\mathcal{Z}_{\phi} = \frac{1}{(2\pi)^{N+1}}~^{M\otimes} \langle \ell=0|\prod_e e^{\beta \cos(\hat\Theta_e-\delta_e\phi)}\ket{\Psi_{0,0}}
\end{equation}
Next, using the fact that
\begin{equation}
	e^{\beta \cos(\hat\Theta_e-\delta_e\phi)}\ket{\ell=0}= \frac{1}{\sqrt{2\pi}}\int_{-\pi}^{\pi} e^{\beta \cos(\Theta_e-\delta_e\phi)}d\Theta_e \ket{\Theta_e}
\end{equation}
and the change of variables \(\Theta'_e=\Theta_e-\delta_e\phi\), one finds that
\begin{equation}
	\mathcal{Z}_{\phi} = \frac{1}{(2\pi)^{N+1}} \bra{\alpha}\prod_e Z_e(\delta_e\phi)\ket{\Psi_{0,0}}
\end{equation}
According to this equation, the twist angle added in the Hamiltonian of the $XY$ model leads to an operator $\prod_e Z_e(\delta_e\phi)$ in the quantum formalism. As shown in Fig.~\ref{twistw}, the twisted boundary is crossed by a non-contractible loop $C_{\bar{y}}$. Therefore, $\prod_e Z_e(\delta_e\phi)$ is the same as non-contractible operator $W_{\bar{y}}(\phi)$ defined in Eq.~(\ref{w}) as a logical operator of toric-rotor code. In particular, according to the definition of the logical states in Eq.~(\ref{wket}) we have $\prod_e Z_e(\delta_e\phi)\ket{\Psi_{0,0}}=\ket{\Psi_{\phi,0}}$. In this regard, we conclude that the partition function of the $XY$ model in the presence of a twist can be expressed as follows:
\begin{equation}
	\mathcal{Z}_{\phi} = \frac{1}{(2\pi)^{N+1}} \bra{\alpha}\ket{\Psi_{\phi,0}}
\end{equation}
This relationship implies that by applying a twist to the $XY$ model, the system is mapped onto another logical state of the toric-rotor code. We again consider the effect of noise on the state \(\ket{\Psi_{0,0}}\) and calculate its fidelity with respect to state \(\ket{\Psi_{\phi,0}}\), as follows:

\begin{equation}
	\mathcal{F}_\phi=\bra{\Psi_{\phi,0}}\mathcal{E}(\rho_0)\ket{\Psi_{\phi,0}}
\end{equation}
This fidelity is equal to the probability that the initial logical state is transformed into another logical state under the effect of the noise.

Now we show that the above fidelity is equal to partition function of $XY$ model with twisted boundary. To this end, we notice that by replacing $|\Psi_{\phi , 0}\rangle$ from Eq.~(\ref{wket}), the fidelity \(\mathcal{F}_\phi\) can be written as
\begin{equation}\label{fphi}
	\mathcal{F}_\phi=\bra{\Psi_{0,0}}\mathcal{E}_{\phi}(\rho_0)\ket{\Psi_{0,0}}
\end{equation}
where \(\mathcal{E}_{\phi}(\rho_0)\) is defined analogously to \(\mathcal{E}(\rho_0)\) in Eq.~(\ref{rho0}), except that the operator \(\prod_e Z_e(\Theta_e)\) is replaced by \(\prod_e Z_e(\Theta_{e}+\delta_e\phi)\). By performing the change of variables \(\Theta'_{e}=\Theta_{e}+\delta_e\phi\), the following expression for \(\mathcal{E}_{\phi}(\rho_0)\) is obtained:
\begin{equation}
	\mathcal{E}_{\phi}(\rho_0)=\int \prod_{e} d\Theta'_{e}\, P_e(\Theta'_{e}-\delta_e\phi) \prod_{e} Z_{e}(\Theta'_{e}) \rho_0
	\prod_eZ_e^{\dagger}(\Theta'_e)
\end{equation}
Then, notice that $\mathcal{F}_\phi$ in Eq.~(\ref{fphi}) has been written in a form similar to Eq.~(\ref{eq00}) and therefore, using a similar argument, it is concluded that \(\mathcal{F}_\phi\) is proportional to the partition function of the \(XY\) model in the presence of a twist (see Eq.~(\ref{patw})). Consequently, one obtains
\begin{equation}
	\mathcal{F}_\phi = \frac{1}{[I_0(1/\sigma)]^M} \mathcal{Z}_\phi
\end{equation}
Now, let us provide a physical interpretation of the results derived from the quantum formalism. In particular, we consider a toric-rotor code initialized in the logical state $|\Psi_{0,0}\rangle$. Then, under the effect of phase-shift noise applied to the rotors, the initial state is converted into a mixture of different logical states and excited states of the code. However, let us limit our attention to the logical subspace of the code, whose density matrix is a mixture of different logical states. Since logical states yield a trivial syndrome when the stabilizers of the code are measured, the above density matrix corresponds to the postselected state conditioned on the zero-syndrome measurement outcome. In particular, according to our results, the probability of each logical state $|\Psi_{\phi,0}\rangle$ in the final mixture is proportional to the partition function of the $XY$ model with twist angle $\phi$. In this regard, $\mathcal{F}_\phi$ is in fact the fidelity between the postselected final state and the initial logical state.

In the next step, we connect the above fidelity relation to the spin stiffness in the $XY$ model. To this end, taking the logarithm of both sides, multiplying by $-T$, and neglecting the constant term, we obtain the following relationship between the logarithm of the fidelity and the free energy $\mathcal{A}_\phi$~\cite{Wang2015}:

\begin{equation}\label{sd}
	-T \ln\mathcal{F}_\phi= \mathcal{A}_\phi.
\end{equation}
Next, we take the second derivative of this equation with respect to the phase parameter \(\phi\):
\begin{equation}
	-\left(\frac{\partial^2 \ln \mathcal{F}_\phi}{\partial \phi^2}\right)_{\phi=0}=\frac{1}{T}\left( \frac{\partial^2\mathcal{A}_\phi}{\partial \phi^2} \right)_{\phi=0}
\end{equation}
According to Eq.~(\ref{rhos2}), the right-hand side of the above equation coincides with the spin stiffness ($\rho_s$) in the $XY$ model, while the left-hand side corresponds to the \textit{fidelity susceptibility}, denoted by $\chi_F$~\cite{Gu2009a}. Consequently, the relationship between the fidelity susceptibility and the spin stiffness is given by:
\begin{equation}
	\chi_F= \frac{L^{d-2}}{T} \rho_s
\end{equation}
In two dimensions ($d=2$), $\chi_F$ is proportional to $\rho_s$ and is independent of the system size. Just as $\rho_s$ measures the stiffness of the classical system with respect to an applied twist, $\chi_F$ characterizes the sensitivity of the logical state in the quantum code to a noise-induced logical gate. In order to clarify this point, let us return to the relation between fidelity and free energy in Eq.~(\ref{sd}). We rewrite this equation as $\mathcal{F}_\phi = \exp(-\frac{\mathcal{A}_\phi}{T})$, so that $\frac{\mathcal{F}_\phi}{\mathcal{F}} = \exp\!\left(-\frac{\mathcal{A}_\phi -\mathcal{A}_0}{T}\right)$. Combining this with the relation between free energy and spin stiffness in Eq.~(\ref{rhos1}), we obtain:
\begin{equation}\label{fphif}
	\mathcal{F}_\phi
	=
	\mathcal{F}
	\exp\!\left(
	-\frac{\rho_s}{2 T}\,
	\phi^2 L^{d-2}
	\right).
\end{equation}
Using the above relation, we can determine the different phases of the noisy toric-rotor code. First, we consider the high-noise regime, where the noise width is greater than $\sigma_c \approx 0.89$, or equivalently the temperature is greater than $T_c$. In this phase, $\rho_s = 0$, and therefore the fidelity $\mathcal{F}_\phi$ is equal to $\mathcal{F}$ for all values of $\phi$. We recall that we consider the postselected density matrix of the system conditioned on the zero-syndrome measurement outcome, where the fidelity $\mathcal{F}_\phi$ corresponds to the probability that the noise effectively implements a logical gate $W_{\bar{y}}(\phi)$. In other words, $\mathcal{F}_\phi$ corresponds to the probability of a logical error occurring in the logical subspace of the code. Therefore, it can be concluded that in the high-noise phase, the toric-rotor mixed state populates the different logical states $|\Psi_{\phi,0}\rangle$ with equal probability. This implies that complete decoherence occurs within the logical subspace when the noise width is greater than $\sigma_c$. Conversely, in the low-noise phase, where $\sigma < \sigma_c$, since $\rho_s \neq 0$, $\mathcal{F}_\phi$ is a Gaussian function of $\phi$ that peaks at $\phi = 0$ and decays as $\phi$ increases. This implies that in the low-noise phase, the logical sectors occur with different probabilities and the system retains a finite degree of coherence in the logical subspace. In other words, the KT phase in the $XY$ model corresponds to a partially coherent phase for the logical subspace of the toric-rotor code, while the disordered phase corresponds to a fully decoherent phase.

\begin{figure}[h!]  
	\centering
	\includegraphics[width=7.1cm,height=4.8cm,angle=0]{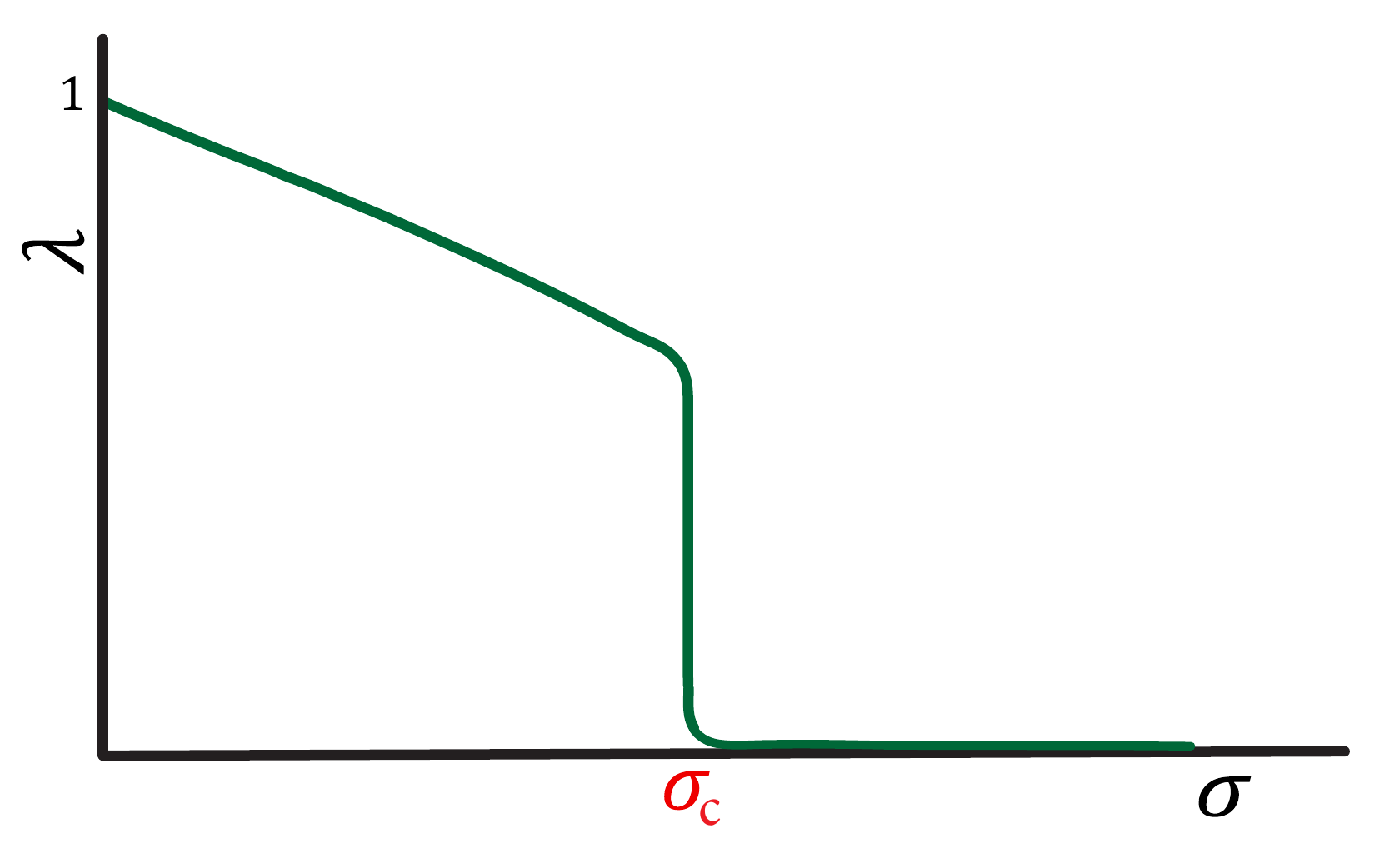} 
	\caption{Resilience order parameter $\lambda$ as a function of the noise width $\sigma$. In the partially coherent phase ($\sigma < \sigma_c$), $\lambda$ satisfies $\lambda < 1$, indicating partial resilience of the logical subspace against noise. At the critical point $\sigma_c \approx 0.89$, the system undergoes a resilience phase transition in which $\lambda$ drops discontinuously to zero, signaling the onset of the fully decoherent phase in which the topological sectors become completely mixed.}
	\label{P7}
\end{figure}

It is also important to characterize this phase transition using a suitable order parameter. To this end, we define the expectation value $\langle \cos \phi \rangle$, evaluated with respect to the probability distribution function $\mathcal{F}_\phi / \int \mathcal{F}_\phi \, d\phi$:
\begin{equation}
	\lambda=\frac{\int_{-\pi}^{\pi} \cos(\phi) e^{\frac{-\rho_s}{2 T} \phi^{2}} \, d\phi}{\int_{-\pi}^{\pi}  e^{\frac{-\rho_s}{2 T} \phi^{2}} \, d\phi}.
\end{equation}
Notice that $\lambda=0$ corresponds to a uniform probability distribution which implies a complete decoherence in the logical subspace of the code while $\lambda=1$ corresponds to a single-value probability distribution $\delta_{\phi ,0}$ which implies a full coherence. On the other hand, according to our argument about the behavior of the probability distribution in the two phases, it is expected that $\lambda$ vanishes in the fully decoherent phase, while it takes a nonzero value in the partially coherent phase; it therefore serves as a suitable order parameter. As shown in Fig.~\ref{P7}, we plot this order parameter using the values of $\rho_s$ obtained from Fig.~\ref{P5}. It shows that $\lambda$ satisfies $0< \lambda < 1$ throughout the partially coherent phase and then drops discontinuously to zero at the critical point $\sigma_c$. In other words, in the partially coherent phase, the topological sectors retain partial resilience to noise, while in the fully decoherent phase, they become completely mixed. Accordingly, we refer to the above phase transition as a resilience phase transition, and we call $\lambda$ a resilience order parameter. We further recall Eq.~(\ref{wket}), in which $e^{-i\phi}$ is the eigenvalue of the logical operator $T_x$ corresponding to $|\Psi_{\phi,0}\rangle$. In this regard, $\lambda$ is also equal to the expectation value of the logical operator $\frac{T_x + T_x^\dagger}{2}$, evaluated with respect to the zero-syndrome postselected density matrix. In particular, the non-locality of this operator implies that the nature of the phase transition is topological. In other words, we have found a correspondence between classical and quantum topological order, in which a classical topological order parameter (the spin stiffness) is mapped to a quantum topological order parameter.

\section{Connection to postselected quantum error correction}\label{sec5}

 It is also important to clarify the difference between the nature of the above resilience phase transition and conventional correctable-to-uncorrectable phase transitions. In particular, notice that to properly address correctability, one should study all postselected ensembles corresponding to different measurement outcomes of the syndrome, whereas here we have limited our study to a specific postselected ensemble, namely the zero-syndrome sector. In this regard, our study is related to a fully postselected error correction, analogous to the framework recently developed for the qubit toric code \cite{English2025a,English2025b}. Moreover, we notice that in our model the value of the resilience order parameter $\lambda$ in the partially coherent phase is not equal to $1$. $\lambda \neq 1$ means that even when the measurement outcome shows no syndrome, the postselected state is not $|\Psi_{0,0}\rangle$ but is instead mixed with other logical states $|\Psi_{\phi,0}\rangle$. In other words, the postselected quantum state is uncorrectable in both the partially coherent and fully decoherent phases of the model. It means that even a very small noise width leads to a logical error in the fully postselected subspace.

On the other hand, for the toric-rotor code in higher dimensions $d>2$, correctability in the fully postselected sector remains possible. In particular, as can be seen in Eq.~(\ref{fphif}), there is a term proportional to $L^{d-2}$ in the exponential. Accordingly, for $d>2$, in the thermodynamic limit $L \to \infty$, $\mathcal{F}_\phi$ approaches zero for any finite value of $\phi$, whereas it remains nonzero only for $\phi = 0$. This implies that there is a completely coherent phase in which the resilience order parameter is equal to $1$, and the code is therefore correctable in the fully postselected sector. In this regard, the critical width $\sigma_c$ for higher-dimensional toric-rotor codes can be regarded as a postselected threshold, in the same sense as above: it marks the noise strength below which $\lambda = 1$ and the fully postselected ensemble is entirely free of logical errors. This threshold is determined solely by the fidelity structure of the noisy code, independently of any particular decoding algorithm, and should therefore be understood as an intrinsic property of the phase structure rather than as the success probability of a specific decoder.

Our results about fully postselected error correction can shed light on conventional (non-postselected) error correction in toric-rotor code. As shown in \cite{English2025a}, for qubit version of toric code, while conventional error correction is mapped to random-bond Ising model, fully postselected error correction is mapped to a clean Ising model. Building on the above mappings, it is shown that the postselected threshold provides an upper bound on the conventional error threshold. In this regard, we conjecture that a relation analogous to the qubit case holds for the toric-rotor code. In particular, our result that no postselected threshold exists for $d=2$ suggests that the toric-rotor code is also uncorrectable in the non-postselected scenario in two dimensions. Conversely, for $d>2$, the existence of a nonzero postselected threshold makes it reasonable to expect a conventional threshold as well, below which the code is correctable. This expectation is consistent with the discussion in Ref.~\cite{Vuillot2024}, where the authors anticipate that the calculation of the conventional threshold maps to a disordered version of the $XY$ model.

Finally, note that, as discussed in the definition of the normalized logical state in Eq.~(\ref{ketphys}), the focus of our study was on the limit $\Delta \rightarrow 0$. In this regard, it is also interesting to consider the effect of the regularization parameter on correctability. In particular, according to Ref.~\cite{Xu2024b}, the situation is expected to be worse in the regularized case. Consequently, for $d = 2$, we anticipate that there is no correctable phase, similar to the case $\Delta \rightarrow 0$. However, for higher dimensions, where a correctable phase in the fully postselected sector exists in the limit $\Delta \rightarrow 0$, we expect that an increase in the regularization parameter leads to a reduction in the value of the critical width $\sigma_c$.

\section{Conclusion}

While quantum formalism for the partition functions of classical spin models have been extensively studied over the past decades, classical models with continuous variables have received less attention. Here, we introduced a quantum formalism for the partition function of the classical $XY$ model and showed that it leads to new insights into the toric-rotor code in the presence of phase-shift noise. In particular, we characterized a resilience phase transition in the zero-syndrome postselected subspace of the noisy toric-rotor code, corresponding to the KT transition of the $XY$ model. We showed that this phase transition occurs as the width of the noise increases, in the sense that the noise width plays a role analogous to temperature in the $XY$ model. We further demonstrated that the spin stiffness of the $XY$ model is related to the resilience of the toric-rotor code to noise. Using well-known properties of the spin stiffness, we identified a resilience order parameter $\lambda$ for characterizing the phase transition in the noisy toric-rotor code.

For the two-dimensional toric-rotor code, we found that $\lambda < 1$ for any nonzero noise width, so that the postselected ensemble remains uncorrectable throughout both the partially coherent and fully decoherent phases; the transition we identify therefore separates two uncorrectable regimes rather than marking a transition in correctability. In contrast, our results suggest enhanced intrinsic resilience to noise for the toric-rotor code in higher dimensions. In particular, the value of the noise width at the transition point for $d>2$ can be regarded as a postselected threshold below which high-dimensional toric-rotor codes are correctable in the fully postselected sector. In this regard, the approach introduced in this paper can shed new light on correctability in continuous-variable quantum codes. However, a complete study of correctability for the higher-dimensional toric-rotor code would require considering the non-postselected scenario. We expect that this problem maps to a random-phase $XY$ model, which we leave for future work in $d>2$. We finally note that our results establish a mapping between classical and quantum versions of topological phase transitions. This can lead to an interesting cross-fertilization between classical and quantum topological order, in the sense that our knowledge of topological order in one setting leads to new insights into the other.

\section*{Acknowledgment}
We would like to thank Mohammad Nobakht for valuable discussions during several meetings that we had on this work.

\bibliography{references} 
 
\end{document}